\documentclass[twocolumn,prl,showpacs]{revtex4}%
\usepackage{graphicx}
\usepackage{bm}
\def\d{{\partial}}
\def\s{{\sigma}}
\def\e{{\epsilon}}
\def\k{{ {\bm k} }}

\def\w{{\omega}}
\def\a{{\alpha}}
\def\b{{\beta}}

\begin{document}

\def\runtitle{
Anomalous Hall Effect in $t_{2g}$ Orbital Kagome Lattice 
due to Non-collinearity: \\
Significance of Orbital Aharonov-Bohm Effect
}
\def\runauthor{
Takeshi {\sc Tomizawa}, and Hiroshi {\sc Kontani}
}

\title{
Anomalous Hall Effect in $t_{2g}$ Orbital Kagome Lattice 
due to Non-collinearity: \\
Significance of Orbital Aharonov-Bohm Effect
}

\author{
Takeshi {\sc Tomizawa}, and Hiroshi {\sc Kontani}
}

\address{
Department of Physics, Nagoya University,
Furo-cho, Nagoya 464-8602, Japan. 
}

\date{\today}

\begin{abstract}
A new mechanism of spin structure-driven anomalous Hall effect (AHE) 
in tilted ferromagnetic metals
is proposed by taking account of the $d$-orbital degree of freedom.
We find that a conduction electron acquires a Berry phase
due to the complex $d$-orbital wavefunction, 
in the presence of non-collinear spin structure and the spin orbit interaction.
The AHE driven by this orbital-derived Berry phase
is much larger than the AHE induced by spin chirality, and
it naturally explains the salient features of 
spin structure-driven AHE in pyrochlore Nd$_2$Mo$_2$O$_7$.
Since the proposed AHE can occur even for coplanar spin orders
($M_z=0$), it is expected to emerge in other 
interesting geometrically frustrated systems.
\end{abstract}

\sloppy

\pacs{72.10.-d, 72.80.Ga, 72.25.Ba}

\maketitle



Recently, intrinsic anomalous Hall effect (AHE) attracts renewed attention
as an interesting quantum transport phenomenon in multiband metals.
It is independent of the quasiparticle damping rate $\gamma$ 
as shown by Karplus and Luttinger (KL) \cite{KL}.
Recently, the theory of KL had been developed intensively, and 
quantitative studies had been performed for $f$-electron systems
 \cite{Kontani94}, $d$-electron systems \cite{Fe,Kontani08},
and two-dimensional Rashba electron gas model 
\cite{Inoue-AHE,Sinova}.
The origin of giant AHE realized in $d(f)$ electron systems
is the ``orbital Aharonov-Bohm (AB) effect'',
in which the complex phase factor arises from the $d(f)$ angular
momentum that is revived by the strong atomic spin-orbit interaction (SOI)
 \cite{Kontani94,Kontani08,Kontani-Ru,Tanaka-4d5d,Kontani-OHE,Wu}.
In $d(f)$ electron systems, strong entanglement of orbital and 
spin degree of freedom in multiband Bloch function 
gives rise to the prominent Berry curvature \cite{Niu,MOnoda}.

\begin{figure}[ptbh]
\includegraphics[width=.99\linewidth]{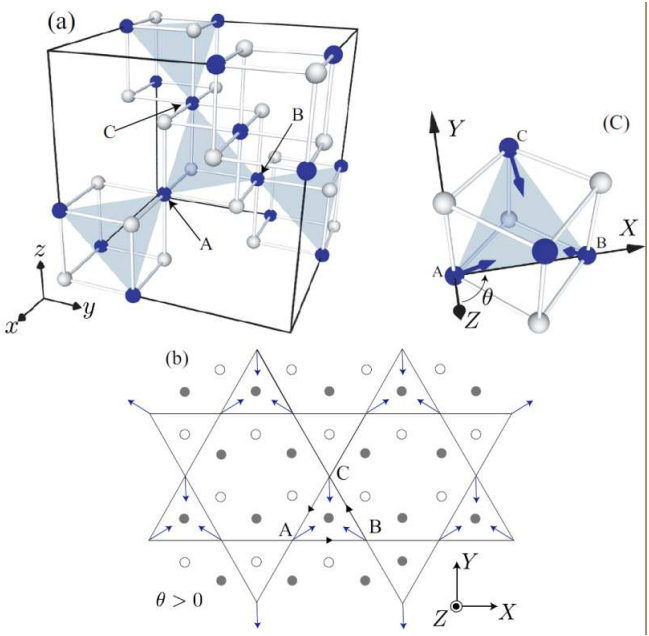}
\caption{(color online)
(a) Spinel crystal structure. Blue (white) circles are Mo (O) ions.
The Mo ions on the [111] plane form the kagome lattice.
(b) Kagome lattice made of Mo ions. 
White (gray) circles are O ions above (below) the Mo layer.
A unit cell contains sites A,B,C.
(c) Umbrella like local exchange field at Mo sites 
represented by arrows.
}
\label{fig:lattice}
\end{figure}

Especially, the AHE due to nontrivial spin structure 
attracts increasing
attention, in accordance with the recent development 
of the study of frustrated systems.
The AHE induced by the Berry phase associated with spin chirality 
had been discussed for Mn-oxides \cite{Millis},
pyrochlore compounds (kagome lattice) \cite{Ohgushi,Lac},
and in spin glass systems \cite{Tatara}.
Metallic pyrochlore Nd$_2$Mo$_2$O$_7$ would be 
the most famous experimental candidate
 \cite{Yoshii,Yasui,Taguchi,Naka}:
Below $T_{\rm c}=93$ K, Mo $4d$ electrons is in
the ferromagnetic state. 
Below $T_{\rm N}\approx 30$ K, localized Nd 4$f$ electrons
form non-coplanar spin-ice magnetic order,
and the tilting of Mo moment $\theta$ is induced
by $d$-$f$ exchange interaction, as shown in Fig. \ref{fig:lattice}.
The chirality-driven anomalous Hall conductivity (AHC) is proportional to 
${\bm s}_{\rm A}\cdot({\bm s}_{\rm B}\times{\bm s}_{\rm C}) \propto \theta^2$
in the weak exchange coupling near half filling
 \cite{Ohgushi,Lac}.
However, 
the neutron diffraction experiments
\cite{Yasui} had shown that $\theta^2\ll 10^{-3}$, suggesting that the
chirality-driven AHC is very small.
Moreover, the field dependences of the AHC and $\theta^2$
are quite different \cite{Yasui}.

In previous studies \cite{Millis,Ohgushi,Lac},
$s$-orbital models had been studied.
However, it is natural to expect a significant role of the 
orbital degree of freedom on the spin structure-driven AHE,
as in the case of the conventional KL-type AHE in $d$-electron systems
 \cite{Kontani94,Kontani08,Kontani-Ru,Tanaka-4d5d,Kontani-OHE}.

In this paper, we find
that the Berry phase is induced by the 
complex $d$-orbital wavefunction in tilted ferromagnetic metals, 
and it causes a prominent spin structure-driven AHE.
In Nd$_2$Mo$_2$O$_7$, the AHE due to orbital Berry phase 
is much larger than the chirality driven AHE, since the former
is {\it linear in $\theta$} consistently with experiments.
In contrast, both the conventional KL term 
in simple ferromagnets ($\propto M_z \propto \cos\theta$)
and the spin chirality term cannot have $\theta$-linear terms.
The present orbital mechanism will be realized 
not only in other pyrochlore Pr$_2$Ir$_2$O$_7$ \cite{Naka},
but also in various geometrically frustrated metals.

Here, we construct the $t_{2g}$ tight-binding model for Mo $4d$ electrons.
For this purpose, we consider the spinel structure (XMo$_2$O$_4$) 
instead of pyrochlore structure (X$_2$Mo$_2$O$_7$).
In both structures, Mo atoms form the same pyrochlore lattice.
The location of O atoms in spinel is rather easy to treat theoretically,
and the difference is not essential for later study,
as we discuss later.
Figure \ref{fig:lattice} (a) represents the Mo atoms (black circles) 
and O atoms (white circles) in the spinel structure.
The [111] Mo layer and the surrounding O atoms in the spinel structure
is extracted in Fig. \ref{fig:lattice} (b), where the white (gray) circles 
represent the O atoms above (below) the Mo layer.
We study this kagome lattice $t_{2g}$-orbital tight binding model, 
where the principal axis for the Mo $d$-orbital is fixed by
the surrounding O$_6$ octahedron.
A vector $(n_x,n_y,n_z)$ in the $xyz$-coordinate in Fig. \ref{fig:lattice} (a)
is transformed into $[n_X,n_Y,n_Z]$ in the $XYZ$-coordinate in 
Fig. \ref{fig:lattice} (b) as $(n_x,n_y,n_z)= [n_X,n_Y,n_Z] {\hat O}$,
where the coordinate transform matrix ${\hat O}$ is given by
\begin{eqnarray}
{\hat O}= \frac{1}{\sqrt{6}}
\left(
\begin{array}[c]{ccc}
-\sqrt{3} & \sqrt{3} & 0 \\
-1 & -1 & 2 \\
\sqrt{2} & \sqrt{2} & \sqrt{2}\\
\end{array}
\right) ,
 \label{eqn:O}
\end{eqnarray}
Arrows in Fig. \ref{fig:lattice} (c) represents the 
local effective magnetic (exchange) field at Mo sites, 
which is composed of the ferromagnetic exchange field from Mo 3$d$ electrons
and the exchange field from Nd 4$f$ electrons.
Under the magnetic field parallel to (1,1,1) direction,
below the N\'{e}el temperature of Nd sites,
the directions of the local exchange fields at sites A, B and C 
in the $XYZ$-coordinate are $(\phi_{\rm A}=\pi/6,\theta)$, 
$(\phi_{\rm B}=5\pi/6,\theta)$ and $(\phi_{\rm C}=3\pi/2,\theta)$, 
respectively \cite{Yoshii,Yasui,Taguchi}.
The tilting angle $\theta$ changes from $-1.5^\circ$ ($H\rightarrow+0$ T)
to $+1.5^\circ$ ($H\sim 6$ T) in Nd$_2$Mo$_2$O$_7$,
corresponding to the change in the spin-ice state at Nd sites
 \cite{Yasui,Sato-private}.

Here, we present a discussion of symmetry.
If we rotate the lattice by $\pi$ around the site C
in Fig. \ref{fig:lattice} (b), the local exchange field is changed 
from $\theta=\theta_0>0$ to $-\theta_0$.
Since the conductivity tensor is unchanged by the $\pi$ rotation,
the AHC due to spin chirality mechanism
is an even function of $\theta$ \cite{Ohgushi,Lac}.
However, O sites in Fig. \ref{fig:lattice} (b) are changed 
by the $\pi$ rotation around C, which means that
the Mo $3d$ orbital state is changed.
For this reason, the AHC in pyrochlore compounds
due to orbital AB phase can have a $\theta$-linear term.

The $t_{2g}$-orbital kagome lattice model is given by
\begin{eqnarray}
H&=&\sum_{i\a,j\b,\s}t_{i\a,j\b}c_{i\a,\s}^\dagger c_{j\b,\s}
-\sum_{i\a,\s\s'} {\bm h}_i \cdot [{\bm \mu}_e]_{\s,\s'} 
c_{i\a,\s}^\dagger c_{i\a,\s'}
 \nonumber \\
& &+\lambda \sum_{i\a\b,\s\s'} [{\bm l}]_{\a,\b} \cdot [{\bm s}]_{\s,\s'} 
c_{i\a,\s}^\dagger c_{i\b,\s'}
 \label{eqn:ham}
\end{eqnarray}
where $i,j$ represent the sites, $\a,\b$ represent the
$t_{2g}$-orbitals ($xy,yz,zx$), and $\s,\s'=\pm1$.
$t_{i\a,j\b}$ is the hopping integrals between $(i,\a)$ and $(j,\b)$.
The direct $d$-$d$ hopping integrals are given by the Slater-Koster (SK)
parameters $(dd\sigma)$, $(dd\pi)$ and $(dd\delta)$.
The third term in eq. (\ref{eqn:ham}) represents the SOI,
where $\lambda$ is the coupling constant, 
and ${\bm l}$ and ${\bm s}$ are the spin and $d$-orbital operators:
Their matrix elements are given in Ref. \cite{Tanaka-4d5d}.
For convenience in calculating the AHC,
we take the $Z$-axis for the spin quantization axis, where
$[s_X,s_Y,s_Z]\times2$ becomes the Pauli matrix vector.
Then, $(s_x,s_y,s_z)= [s_X,s_Y,s_Z]{\hat O}$.
The second term in eq. (\ref{eqn:ham}) represents the Zeeman term, where 
${\bm h}_i$ is the local exchange field at site $i$.
${\bm \mu}_e \equiv -2{\bm s}$ is the magnetic moment of an electron.

The Green function is given by a $18\times18$ matrix:
${\hat G}_\k(\e)=((\e+\mu){\hat 1}-{\hat H}_\k)^{-1}$,
where $\mu$ is the chemical potential and 
${\hat H}_\k$ is the matrix for the Hamiltonian 
in the momentum space.
According to the linear response theory, the AHC is given by 
$\s_{\rm AH}= \s_{\rm AH}^I + \s_{\rm AH}^{I\!I}$ 
\cite{Streda}:
\begin{eqnarray}
\s_{\rm AH}^{I} &=& \frac{1}{2\pi N}\sum_{\k}
{\rm Tr}\left[{\hat j}_X {\hat G}^R {\hat j}_Y {\hat G}^A 
\right]_{\e=0},  \label{eqn:AHCI}
 \\
\s_{\rm AH}^{I\!I} &=& \frac{-1}{4\pi N} \sum_{\k}\int_{-\infty}^0 d\e
{\rm Tr}\left[{\hat j}_X \frac{{\d{\hat G}}^R}{\d\e} 
{\hat j}_Y {\hat G}^R \right.
 \nonumber \\
& &\left. \ \ \ \ \ \ \ \ 
 -{\hat j}_X {\hat G}^R
{\hat j}_Y \frac{{\d{\hat G}}^R}{\d\e} 
- \langle {\rm R}\rightarrow {\rm A} \rangle
\right].
 \label{eqn:AHCII}
\end{eqnarray}
where ${\hat G}_\k^{R(A)}(\e)\equiv {\hat G}_\k(\e+(-)i\gamma)$
is the retarded (advanced) Green function.
${j}_{\k \mu} \equiv -e \d {\hat H}_\k/\d k_\mu$ ($\mu=X, Y$)
is the charge current.
Since all the matrix elements in ${j}_{\k \mu}$ are odd 
with respect to $\k$, the current vertex correction 
due to local impurities vanishes identically \cite{Kontani08,Tanaka-4d5d}.
In the band-diagonal representation, eqs. (\ref{eqn:AHCI}) and 
(\ref{eqn:AHCII}) are transformed into eqs. (30) and (32)-(33)
in Ref. \cite{Kontani08}.

\begin{figure}[ptbh]
\includegraphics[width=.99\linewidth]{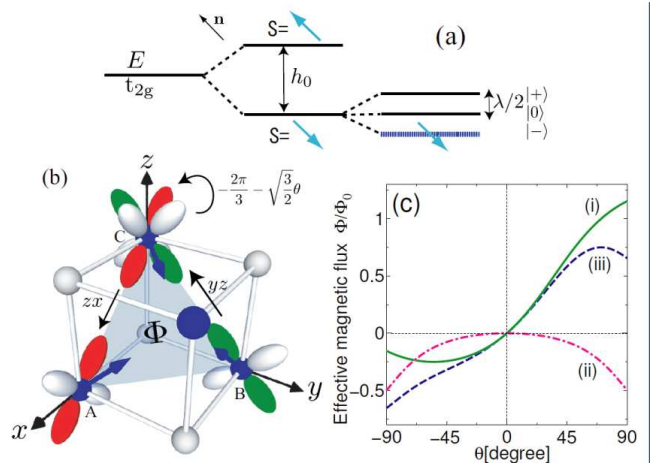}
\caption{(color online)
(a) Eigenenergies for $t_{2g}$ electron under the exchange
field ${\bm h}$ and the SOI $(-\lambda/2){\bm n}\cdot {\bm l}$.
(b) Orbital AB phase given by the complex $t_{\rm 2g}$ orbital 
wavefunction at site C.
(c) Fictitious magnetic flux due to orbital AB effect (i),
due to spin chirality (ii), and the total flux (iii).
}
\label{fig:flux}
\end{figure}

Before proceeding to the numerical results,
we explain an intuitive reason why prominent AHE is induced 
by the non-collinearity of the local exchange field ${\bm h}_i$.
For this purpose, we assume the strong coupling limit where the 
Zeeman energy is much larger than Fermi energy $E_{\rm F}$ and the SOI.
Since ${\bm \mu}_e = -2{\bm s}$,
the SOI term at site $i$ is replaced with $(-\lambda/2){\bm l}\cdot{\bm n}_i$,
where ${\bm n}_i \equiv {\bm h}_i/|{\bm h}_i|$.
Its eigenenergies in the $t_{2g}$ space are $0$ and $\pm \lambda/2$,
as shown in Fig. \ref{fig:flux} (a).
The eigenstate for $E=-\lambda/2$ is given by
\begin{eqnarray}
|{\bm n}\rangle &=& \frac{1}{\sqrt{2(n_x^2+n_y^2)}}
 \large[ -(n_x n_z -i n_y)|xy\rangle
 \nonumber \\
 & &+(n_y^2+n_z^2)|yz\rangle -(n_x n_y +in_z) |zx\rangle \large]
\label{eqn:GS}
\end{eqnarray}
where ${\bm n}=(n_x,n_y,n_z)$ in the $xyz$-coordinate:
${\bm n}_\Xi$ ($\Xi=$A,B,C) in the $xyz$-coordinate is given by
$[\sin\theta\cos\phi_\Xi,
\sin\theta\cos\phi_\Xi, \cos\theta]{\hat O}$.
When $\theta=0$ ($n_{x,y,z}=1/\sqrt{3}$),
eq. (\ref{eqn:GS}) becomes
$[ \w |xy\rangle + |yz\rangle + \w^* |zx\rangle ]/\sqrt{3}$,
where $\w=\exp(i2\pi/3)$.
Considering that $|\theta|\sim0.01$ in Nd$_2$Mo$_2$O$_7$ \cite{Yasui},
we expand eq. (\ref{eqn:GS}) up to the first order in $\theta$.
For site C, it is given as
$|{\rm C}\rangle =  [ 
 (1+\frac{\theta}{\sqrt{2}} )
 \w e^{-i\frac{1}{2}\sqrt{\frac{3}{2}}\theta} |xy\rangle 
+ (1-\frac{\theta}{2\sqrt{2}} ) |yz\rangle 
+ (1-\frac{\theta}{2\sqrt{2}} )
 \w^* e^{-i\sqrt{\frac{3}{2}}\theta} |zx\rangle ]/\sqrt{3}$.
TABLE \ref{table:phase} shows the phases for $|\a\rangle$ orbital 
at site $\Xi$ in the eigenstate in eq. (\ref{eqn:GS}), $\psi_\a^\Xi$, 
up to $O(\theta)$.

\begin{table}[!htbp]
\caption{\label{table:phase} 
Phases for $t_{2g}$ orbitals in eq. (\ref{eqn:GS})
up to $O(\theta)$.
}
\begin{ruledtabular}
\begin{tabular}{c||c|c|c}
  & $\psi_{xy}^\Xi$ & $\psi_{yz}^\Xi$ & $\psi_{zx}^\Xi$ \\ \hline
$\Xi=$A & $\frac{2\pi}{3}-\frac12\sqrt{\frac32}\theta$ & 0 & $-\frac{2\pi}{3}+\frac12\sqrt{\frac32}\theta$ \\ 
$\Xi=$B & $\frac{2\pi}{3}+\sqrt{\frac32}\theta$ & 0 & $-\frac{2\pi}{3}+\frac12\sqrt{\frac32}\theta$ \\
$\Xi=$C & $\frac{2\pi}{3}-\frac12\sqrt{\frac32}\theta$ & 0 & $-\frac{2\pi}{3}-\sqrt{\frac32}\theta$ \\ 
\end{tabular}
\end{ruledtabular}
\end{table}


Here, we explain that a moving electron acquires the ``orbital AB phase'',
which gives rise to the prominent AHE
that is sensitively controlled by the tilting angle $\theta$.
Figure \ref{fig:flux} (b) shows 
the motion of an electron that enter into site C 
via $yz$ orbital and exit via $zx$ orbital.
When the electron is in the eigenstate $|{\rm C}\rangle$,
the electron acquires the phase difference between $yz$ and $zx$ orbitals,
$\exp(i(\psi_{zx}^{\rm C}-\psi_{yz}^{\rm C}))=\w^* e^{-i\sqrt{3/2}\theta}$,
which is the orbital AB phase that is controlled by $\theta$.
The total orbital AB phase factor for the triangle path along 
A$\rightarrow$B$\rightarrow$C$\rightarrow$A in Fig. \ref{fig:flux} (b) 
is given by the phase of the following amplitude:
\begin{eqnarray}
T_{\rm orb}=\langle {\rm A}|H_0|{\rm C}\rangle\langle {\rm C}|H_0|{\rm B}\rangle\langle {\rm B}|H_0|{\rm A}\rangle
 \label{eqn:Torb}
\end{eqnarray}
where $H_0$ is the kinetic term. 
Here, we take only the most largest SK parameter $(dd\sigma)$,
and neglect $(dd\pi)$ and $(dd\delta)$ to simplify the discussion.
In this case, only the following intraorbital hoppings exist:
$\langle {\rm B};xy|H_0|{\rm A};xy \rangle
= \langle {\rm C};yz|H_0|{\rm B};yz \rangle
=\langle {\rm A};zx|H_0|{\rm C};zx \rangle = t$.
By concentrating only on the phase given in TABLE \ref{table:phase}, 
we obtain
\begin{eqnarray}
T_{\rm orb}&\sim& (t^3/27) e^{i(\psi_{zx}^{\rm C}-\psi_{yz}^{\rm C})
+i(\psi_{yz}^{\rm B}-\psi_{xy}^{\rm B})
+i(\psi_{xy}^{\rm A}-\psi_{zx}^{\rm A})}
 \nonumber \\
&=& (t^3/27) \exp(-i3\sqrt{3/2}\theta) +O(\theta^2).
 \label{eqn:Torb2}
\end{eqnarray}
$T_{\rm orb}$ is also expressed as 
$T_{\rm orb}=|T_{\rm orb}|e^{-i2\pi\Phi_{\rm orb}/\Phi_0}$, 
where $\Phi_0=2\pi\hbar/|e|$ is the flux quantum, and $\Phi_{\rm orb}$ is the 
``fictitious magnetic flux'' due to the orbital AB effect.
Therefore, $\Phi_{\rm orb}=(3/2\pi)\sqrt{3/2}\theta\cdot \Phi_0$
up to $O(\theta)$.
Since the relation $\Phi_{\rm orb}\propto \theta$ also holds
in pyrochlore compounds,
the orbital AB flux should give rise to the AHE 
that is linear in $\theta$ in Nd$_2$Mo$_2$O$_7$.
The line (i) in Fig. \ref{fig:flux} (c) represents $\Phi_{\rm orb}$
for for $|\theta|\le \pi/2$, which is given by using eq. (\ref{eqn:Torb})
and the relationship $-(\Phi_0/2\pi){\rm Im}\ln(T_{\rm orb}/|T_{\rm orb}|)$.
At $\theta=\pm \pi/2$, the obtained $\Phi_{\rm orb}$ 
is not an integer multiple of $\Phi_0/2$.
This fact means that the AHC is finite even in the case of coplanar order.

In the above discussion, we have neglected interorbital hoppings integrals
for simplicity.
If we include them, the AHC is finite even if $\theta=0$ since the 
orbital AB phase is nonzero, as discussed in Refs. \cite{Kontani08,Kontani-Ru}.
Thus, interorbital hoppings integrals are necessary to realize the AHE
for $\theta=0$ (conventional KL type AHE).

We also discuss the fictitious magnetic flux due to the spin chirality,
which comes from the Berry phase accompanied with
the spin rotation due to the electron hopping
\cite{Ohgushi}.
The line (ii) in Fig. \ref{fig:flux} (c) represents
the fictitious magnetic flux due to the spin chirality 
$\Phi_{\rm spin}$:
Geometrically, $-4\pi\Phi_{\rm spin}$ corresponds to the solid angle
subtended by ${\bm n}_{\rm A}$, ${\bm n}_{\rm B}$ and ${\bm n}_{\rm C}$.
Since $\Phi_{\rm spin} \propto \theta^2$ for $\theta\ll1$,
spin chirality mechanism is negligible in Nd$_2$Mo$_2$O$_7$ \cite{Yasui}.
The line (iii) shows the total flux 
$\Phi_{\rm tot}=\Phi_{\rm orb}+\Phi_{\rm spin}$.

Thus, all the upward and downward ABC triangles 
in Fig. \ref{fig:lattice} (b) are penetrated by $\Phi_{\rm tot}$.
However, the total flux in the unit cell cancels out
since the hexagons are penetrated by $-2\Phi_{\rm tot}$.
Nonetheless, the AHC becomes finite  \cite{Ohgushi,Haldane}
since the contribution from the triangle path, which is the third order of 
the hopping integrals, would be the most important.
The main features of the AHE due to orbital AB effect,
$\s_{\rm AH}^{\rm orb}$, are the following:
(a) $\s_{\rm AH}^{\rm orb}$ dominates the spin chirality AHE for $\theta\ll1$.
(b) $\s_{\rm AH}^{\rm orb}$ exists even if $\theta=\pi/2$ (coplanar order),
whereas spin chirality AHE vanishes 
since $\Phi_{\rm spin}(\theta=\pi/2)=\Phi_0/2$.

\begin{figure}[ptbh]
\includegraphics[width=.99\linewidth]{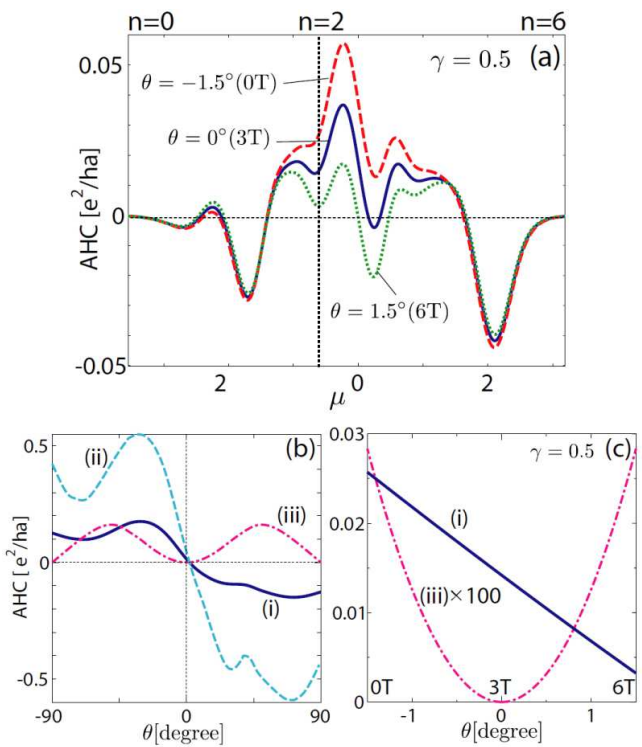}
\caption{(color online)
(a) AHC for $\theta=0$ and $\pm 1.5^\circ$ as function of $\mu$. 
(b) AHC in the present $t_{2g}$-orbital model for (i) $\gamma=0.5$ and 
(ii) $\gamma=0.2$. 
(iii) shows the AHC in the $s$-orbital model for $\gamma=0.5$.
(c) (i) and (iii) for $|\theta|\le1.5^\circ$.
}
\label{fig:calc}
\end{figure}

From now on, we perform numerical calculation for the AHC
using eqs. (\ref{eqn:AHCI}) and (\ref{eqn:AHCII}),
which contains the effects from both $\Phi_{\rm orb}$ and $\Phi_{\rm spin}$,
and verify that the concept of orbital AB effect is still appropriate
even in the weak coupling regime where $|{\bm h}_i|\ll E_{\rm F}$.
Here, we put the SK parameters between the nearest neighbor Mo sites as 
$(dd\s)=-1.0$, $(dd\pi)=0.6$, and $(dd\delta)=-0.1$.
$|(dd\s)|$ corresponds to 2000 K.
The spin-orbit coupling constant for Mo 4$d$ electron is 
$\lambda=0.5$ \cite{Tanaka-4d5d}. 
The number of electrons per site is $n=2$ ($1/3$-filling)
for Nd$_2$Mo$_2$O$_7$ since the valence of Mo ion is $4+$.
To reproduce the magnetization of Mo ion $1.3\mu_{\rm B}$ in Nd$_2$Mo$_2$O$_7$,
we set $|{\bm h}_i|=0.9$.
We also put the damping rate $\gamma=0.2\sim0.5$
to reproduce the resistivity $\sim 1$ m$\Omega$cm
in a single crystal \cite{Taguchi}.

Figure \ref{fig:calc} (a) shows the $\mu$-dependence of the AHC
for $\theta=-1.5^\circ$ ($\sim 0$ T), $\theta=0$ ($\sim3$ T),
and $\theta=1.5^\circ$ ($\sim6$ T).
The AHC for $\theta=0$ is given by the conventional KL mechanism.
Since the present spin structure-driven AHE is linear in $\theta$,
a very small $\theta$ causes a prominent change 
in the AHC for $1.3>\mu>-1.3$ ($4.7>n>1.0$).
Since $e^2/ha=10^3 \Omega^{-1}{\rm cm}^{-1}$ for $a=4$\AA
($a$ is the interlayer spacing),
the AHC for $n=2$ and $\theta=1.5^\circ$ corresponds to 
$30 \ \Omega^{-1}{\rm cm}^{-1}$, which is consistent with experiments.

Here, we analyze the $\theta$-dependence of the AHC in detail,
by ignoring the experimental constraint $|\theta|\le1.5^\circ$.
Figure \ref{fig:calc} (b) shows the AHCs as functions of $\theta$.
The lines (i) and (ii) represent the total AHC 
in the present model for $\gamma=0.5$ and $0.2$, respectively.
Their overall functional forms are approximately odd with respect to $\theta$,
and it takes a large value even if $\theta=\pm\pi/2$ (coplanar order).
These results are consistent with the AHE induced by
the orbital AB effect discussed in Fig. \ref{fig:flux}.
Note that $\gamma>0.2$ corresponds to ``high-resistivity regime'' 
where the intrinsic AHC follows an approximate scaling relation
$\s_{\rm AH}\propto \rho^2$ \cite{Kontani94,Kontani08}.
The line (iii) in Fig. \ref{fig:calc} (b) shows the AHC in the 
$s$-orbital model studied in Ref. \cite{Lac}.
We put the nearest neighbor hopping $t=-0.8$, 
$\gamma=0.5$, and $|{\bm h}|=0.9$.
We also put $n=2.5/3=0.83$ since $n=2/3$ ($1/3$-filling) 
corresponds to band insulating state \cite{Lac}.
Although the obtained AHC takes a large value for $\theta\sim 50^\circ$,
it is 100 times smaller than the AHC in the $t_{2g}$ model (i) 
for $|\theta|\le 1.5^\circ$, as shown in Fig. \ref{fig:calc} (c).

Here, we make comparison with the theory and experiments for Nd$_2$Mo$_2$O$_7$:
Under ${\bm H}\parallel (1,1,1)$ below $T_{\rm N}$, $\s_{\rm AH}$ decreases 
with $H$ monotonically from $H\rightarrow0$ T ($\theta\approx -1.5^\circ$)
to 6 T ($\theta\approx 1.5^\circ$).
This experimental result is consistent with the AHE due to 
the orbital AB effect that is linear in $\theta$.
The present study also reproduces
the experimental phenomenological equation 
$\rho_{\rm H}^{\rm AHE}= 4\pi R_s M_Z^{\rm Mo}+4\pi R_s' M_Z^{\rm Nd}$:
The first term is the conventional KL term ($\theta=0$), and the second term 
is the $\theta$-linear term since
$\theta\propto\sqrt{h_X^2+h_Y^2}$ and $\sqrt{h_X^2+h_Y^2} \propto M_Z^{\rm Nd}$ 
due to noncoplanar magnetic order of Nd moments
\cite{Yasui,Sato-private}.

The realization condition for the AHE in the present $t_{2g}$ model
is much more general than that in the $s$-orbital model.
For example, the AHC is finite even in the coplanar spin state
($M_z=0$), in high contrast to the chirality-driven AHE.
Moreover, according to eq. (\ref{eqn:Torb2}), orbital AB phase emerges
if only one of $\theta_{\rm A}, \theta_{\rm B}, \theta_{\rm C}$ is nonzero.
Therefore, in the present $t_{2g}$ model,
prominent AHE due to non-collinearity is realized unless 
${\bm n}_{\rm A} \parallel {\bm n}_{\rm B} \parallel {\bm n}_{\rm C}$,
with the aid of the orbital AB effect and the SOI.

In summary, we studied the AHE in the $t_{2g}$-orbital model
in the presence of non-collinear magnetic configurations.
Thanks to the SOI, local exchange field modifies the 
complex $d$-orbital wavefunction,
and the resultant Berry phase induces prominent AHE that is
much larger than the chirality-driven AHE.
Since the derived AHC is linear in the tilting 
angle of Mo moment, $\theta$ , experimental large 
field dependence in Nd$_2$Mo$_2$O$_7$ is reproduced.


We are grateful to M. Sato and Y. Yasui for valuable discussions.
This work was supported by Grant-in-Aid for Scientific Research 
on Priority Areas ``Novel States of Matter Induced by Frustration''.
We also acknowledge hospitality of STCM-Kyoto in Yukawa Institute 
in 2008.


\end{document}